\documentclass[preprint,aps,showpacs]{revtex4}
\usepackage{graphicx}
\usepackage{tabularx}

\begin{document}

\title{A Further Measurement of the $\beta$-Delayed $\alpha$-Particle Emission of $^{16}N$}
\thanks{Supported by USDOE grants No. DE-FG02-91ER40609 and DE-FG02-94ER40870}

\author{R.H. France III}
\altaffiliation[Also at: ]{A.W. Wright Nuclear Structure Laboratory, P.O. Box 208124,
Yale University, 272 Whitney Avenue, New Haven, CT 06520-8124}
\affiliation{Department of Chemistry 
\& Physics, Campus Box 82, Georgia College \& State University,
Milledgeville, Georgia 31061}
\author{E.L. Wilds}
\affiliation{
Division of Radiation Safety, Connecticut-DEP, 79 Elm Street, Hartford, CT 06106}
\author{J.E. McDonald}
\affiliation{Department of Physics, University of Hartford, 200 Bloomfield Avenue,
West Hartford, CT 06117-1599} 
\author{M. Gai}
\affiliation{Laboratory for Nuclear Sciences at Avery Point, University of Connecticut, 
1084 Shennecossett Road, Groton, CT 06340-6097. \\
and \\ Department of Physics, WNSL-102, P.O. Box 208124,
Yale University, 272 Whitney Avenue, New Haven, CT 06520-8124.}

\begin{abstract}

We measured the $\beta$-delayed $\alpha$-particle emission spectrum of $^{16}N$ with a sensitivity for
$\beta$-decay branching ratios of the order of $10^{-10}$.
The $^{16}N$ nuclei were produced using the $d(^{15}N,^{16}N)p$
reaction with 70 MeV $^{15}N$ beams and a deuterium gas target
7.5 cm long at a pressure of 1250 torr. The $^{16}N$ nuclei were collected
(over 10 s) using a thin aluminum foil  with an areal density of 180 
$\mu g/cm^2$ tilted at $7^\circ$ with respect to the beam.
The activity was transferred to the counting area by means of a stepping motor
in less than $3\ s$ with the counting carried out over $8\ s$.
The $\beta$-delayed $\alpha$-particles were measured using a time of flight method to achieve 
a sufficiently low background. Standard calibration sources ($^{148}Gd, \ ^{241}Am, \ 
^{208,209}Po, \ and \ ^{227}Ac$) as well as alpha-particles and $^7Li$ from the 
$^{10}B(n,\alpha)^7Li$ reaction were used for an accurate energy calibration.
The energy resolution of the catcher foil  (180-220 keV) was calculated 
and the time of flight resolution (3-10 nsec) was measured using the $\beta$-delayed $\alpha$-particle emission
from $^8Li \ $ that was  produced using the $d(^7Li,^8Li)p$ reaction with the same setup.  
The line shape was corrected to account for the variation in the energy and time 
resolution and a high statistics spectrum of the $\beta$-delayed $\alpha$-particle emission of $^{16}N$ is reported. 
However, our data (as well as earlier Mainz data and unpublished Seattle data) do not agree with
an earlier measurement of the $\beta$-delayed $\alpha$-particle emission of $^{16}N$ taken at TRIUMF after averaging over the
energy resolution of our collection system. This disagreement, amongst other issues, prohibits 
accurate inclusion of the f-wave component in the R-matrix analysis.

\end{abstract}

\preprint{UConn-40870-0036, GCSU-CBX82-0003}

\maketitle

\section{Introduction}

The $^{12}C(\alpha,\gamma)^{16}O$ reaction is of critical importance for
understanding stellar evolution \cite{Fo84}. It competes with the
$^8Be(\alpha,\gamma)^{12}C$ reaction (that forms carbon) to yield oxygen during stellar
helium burning, together determining the carbon/oxygen (C/O) ratio at the end of helium burning.
The C/O ratio is of major importance for understanding type II \cite{We93} and 
 apparently also the light curve of type Ia \cite{Ho02}
supernovae. Since the $^8Be(\alpha,\gamma)^{12}C$ reaction is comparatively well
known ($\pm 12\%$), the $^{12}C(\alpha,\gamma)^{16}O$ reaction
provides the principle uncertainty in the
C/O ratio at the end of helium burning.

The $\beta$-delayed $\alpha$-particle emission of {$^{16}N$} ({\em i.e.} $\alpha$-particles
emitted from the continuum of $^{16}O$ populated by the
$\beta$-decay of {$^{16}N$}) has been predicted
to provide a constraint on the cross section
of this reaction \cite{Ba69,Ba71,Ji90,Hu91}, but it
requires a  measurement of a $\beta$-decay of $^{16}N$ with a sensitivity for 
a Branching Ratio (BR) of the order of $10^{-9}$. In particular, the low energy portion of the
alpha-particle spectrum has been predicted to be sensitive 
to the reduced $\alpha$-particle width of the bound
$1^-$ state in $^{16}O$; however, it cannot directly 
determine the mixing phase in the $^{12}C(\alpha,\gamma)^{16}O$ reaction of the
two interfering states at $7.12\ MeV$ and $9.58\ MeV$ in $^{16}O$.

In the early 1970s F.C. Barker
\cite{Ba69,Ba71} proposed the use of the $\beta$-delayed $\alpha$-particle emission spectrum of
$^{16}N$ to constrain $S_{E1}$, the p-wave component
of the astrophysical cross section factor.
At the time of these calculations \cite{Ba69,Ba71},
a single $\beta$-delayed $\alpha$-particle emission spectrum with very high statistics ( $>36$ million counts)
 existed.  These data were measured at Mainz
\cite{Wa69,Wa70,Ne74} in a successful measurement of the parity
violating $1.281\ MeV$ $\alpha$-particle decay. However, these data were only listed in 
numerical form  in private communications \cite{Barker}.
Unfortunately, this spectrum excluded the energy
region of the interference peak predicted at $1.1\ MeV$.

In the early 1990s three additional measurements of
this spectrum were made: at TRIUMF \cite{Az94},
at Yale \cite{Zh93,Zh93a}, and  at the University of 
Washington at Seattle (unpublished) \cite{Zh95}.
The experiment reported here \cite{Fr96,France} is a continuation and improvement
of the original Yale-UConn experiment \cite{Zh93,Zh93a}. We refer the reader to
the appendix of Ref. \cite{France} where all available data from each of these
experiments, including the unpublished Seattle experiment \cite{Zh95} and the old Mainz 
experiment \cite{Wa69,Wa70,Ne74}, are listed in tabular form.

Recently a renewed interest in the spectrum of the  $\beta$-delayed $\alpha$-particle emission of $^{16}N$ has been generated by a new 
experiment carried out at the Argonne National Lab \cite{Argonne}. The primary purpose of this 
paper is  to publish a detailed account of the data from our improved experiment 
that were already shown in the literature \cite{Fr96}. In addition we include in the appendix 
of this paper the same numerical listing as included (by permission) in Ref. \cite{France}
of the Seattle data \cite{Zh95} as well as the Mainz data \cite{Wa69,Wa70,Ne74}.
The Seattle and Mainz data have been repeatedly discussed by a 
number of authors that quoted Reference \cite{France} as the source for these data, 
and we consider it advantageous to list it here in numerical form. We compare 
these data sets and demonstrate that our data agrees with the Mainz and Seattle data but not 
with the TRIUMF data. We also note that the preliminary reported data of the Argonne 
group \cite{Argonne} agree with our data. We discuss the relevance of this 
disagreement for the global R-matrix fit of the data.

\section{Experimental Procedures}

\begin{figure}
 \includegraphics[height=3in]{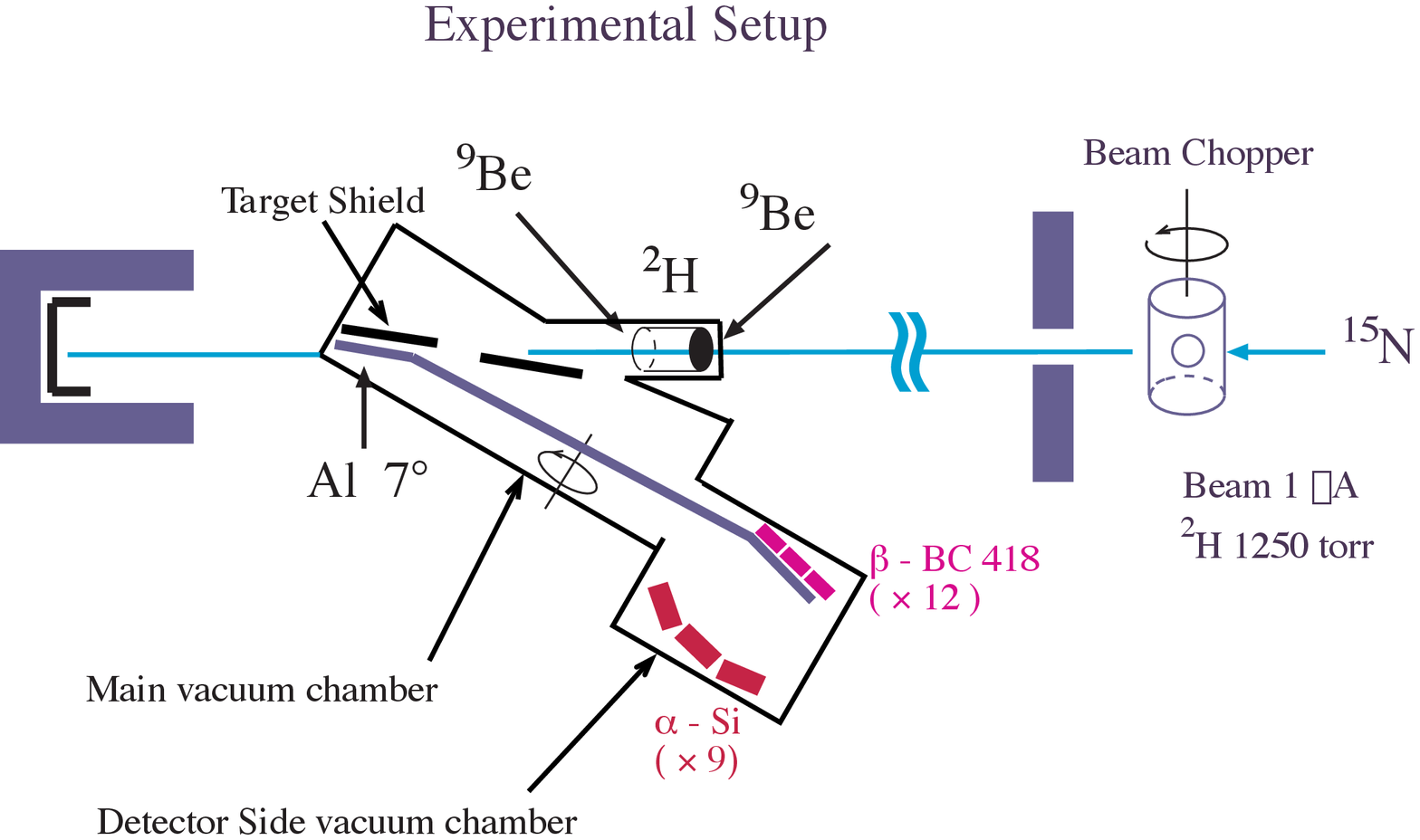}
 \caption{\label{diag} Diagram of the experimental setup 
 showing the rotating arm and the two stations
 (collection and counting).  The beam chopper is far upstream and is well shielded from the 
 experimental hall where the data were collected.  Figure is not drawn to scale.}
\end{figure}

The experiments were performed using $150-250\ pnA$
$70\ MeV$ $^{15}N$ beams from the Yale ESTU tandem
van de Graff accelerator
and a deuterium gas target to produce {$^{16}N$} using the
$d(^{15}N,^{16}N)p$ reaction. The produced $^{16}N$ nuclei emerged
from the gas target and were stopped in aluminum catcher foils, as shown in Fig.~\ref{diag}.
These foils were then rotated into a counting area to
measure the decays of $^{16}N$.  The use of reversed kinematics
allowed for the {$^{16}N$} to be kinematically focussed
into a forward cone of about $7^\circ$.
The {$^{16}N$} nuclei were implanted into thin aluminum catcher foils located at the
ends of an approximately $1\ m$ long arm which rotated
about its center.  After each production cycle, the
arm rotated $180^\circ$, placing the
implanted catcher foil between two detector arrays
which measured the beta and alpha-particles in coincidence.

We used a deuterium gas target, composed of a copper cylinder $7.5\ cm$
long and $1.9\ cm$ in diameter with
$0.3\ cm$ thick walls cooled with an alcohol
cooling system to $-40^\circ C$. Thin beryllium
pressure foils $(25\ \mu m)$ were attached with Araldite
$AW\ 106/HV\ 953$ epoxy from CIBA-GEIGI (good for
very low temperatures)
to the ends of the cylinder.  To reduce plural scattering,
beryllium was chosen for the pressure windows as it has a
low $Z\ (Z=4)$.  The target was filled
with approximately $1250\ torr$ of deuterium gas and 
was placed close to the catcher foils with
its exit window only $7.5\ cm$ from the edge of the catcher foil.

The $^{16}N$ emerged from the target with a broad
distribution of energies (of order $7\ MeV$) of which
the catcher foils collected the lowest $1\ MeV$ portion.  To
allow the capture of these ions while retaining useful
$\alpha$-particle energy resolution, the catcher foils were made of
thin $(180\ \mu g/cm^2)$ aluminum tilted at $7^\circ$ with respect to the
beam, see Fig.~\ref{diag}, to increase their effective catching thickness
by a factor of 8.

The foils were attached to aluminum frames with
epoxy and had open areas of $5\ cm \times 20\ cm$.  Inside the
chamber, see Fig.~\ref{diag}, a
tantalum shield was positioned less than 1 mm from the catcher foil to prevent $^{16}N$
from hitting the catcher frames.   Very precise alignment of the shields
and catcher foils was necessary to ensure that the
$^{16}N$ was stopped in the catcher foils and
not elsewhere, where it would produce a low energy tail
in the alpha-particle spectrum.  The alignment was tested 
using empty catcher foil frames.
The $^{15}N$ beam was confined to travel through the system using
two sets of beam defining slits:
one slit located several meters upstream and the second one about a meter
upstream. The beam position was also constrained by two additional tantalum collimators, 
the target itself, and the shield, discussed above.

During the collection period the neutron background was very large
hence the data acquisition system was turned off during irradiation.
The experiments were run in 21 second cycles, to maximize detection
efficiency given the 10 second lifetime of $^{16}N$.  The first
10 seconds of each cycle was the production
period during which the {$^{16}N$} was produced in the
target and collected in the catcher foils.   At the end
of this time, a tantalum beam chopper blocked the
beam far upstream followed by the arm carrying the catcher
foil rotating $180^\circ$ in
slightly less than $3\ s$.  Three seconds after
the rotation began, the data acquisition system was
activated for $8\ s$.  In order to protect the catcher foil frames
from the beam, the position of the beam
chopper was read back into the control room, and the
arm rotation did not begin until the beam chopper was
fully in place.  Since catcher foils were located at both ends of the arm, it was not necessary to
rotate the arm at the end of the cycle and a second (10 s) collection period commenced.

The principle detectors in this experiment were nine
Silicon Surface Barrier (SSB) detectors located in a
  square array and used to detect $\alpha$-particles.  Each detector had an active
area of $450\ mm^2$ and a $50\ \mu m$ thick active region
to minimize deposition of energy from $\beta$-particles.  Canberra 2003B
preamps, which had been modified to match the high capacitance of the detectors,
 were used.  The  measured energy
resolution of the detectors was about $55\ keV$, and
the array was located $83\ mm$ from the catcher foil.

The secondary detector array (the $\beta$-array) was
composed of twelve plastic scintillation detectors
made of BC418 fast plastic scintillator and
Hamamatsu H3165 and H3171 photomultipliers and was used to
detect $\beta$-particles.  The central
six detectors were $2.5 \times 2.5 \times 0.6\ cm$, while the outer six detectors 
were $5.0 \times 5.0 \times 0.6\ cm$.  The detectors were
optically isolated from one another using aluminum foil and were covered
by a $0.02\ mm$ aluminized mylar film to prevent the detection
of $\alpha$-particles or $^{12}C$ recoils. The $\beta$-array subtended approximately 
30\% of 4$\pi$, as shown in Fig.~\ref{diag}.

The data were collected event by event and written to Exabyte tapes
by a Concurrent 3230 computer running the Oak Ridge data acquistion
system \cite{France}.  Each event was started by one of the $\alpha$-detectors
firing.  For each $\beta$-detector which fired, a delayed relative timing
signal was recorded using a Lecroy 2228A time to digital converter.
  Thus for each event one $\alpha$-particle energy and
12 $\alpha$-$\beta$ relative time measurements were recorded.

\begin{figure}
 \includegraphics[height=4in]{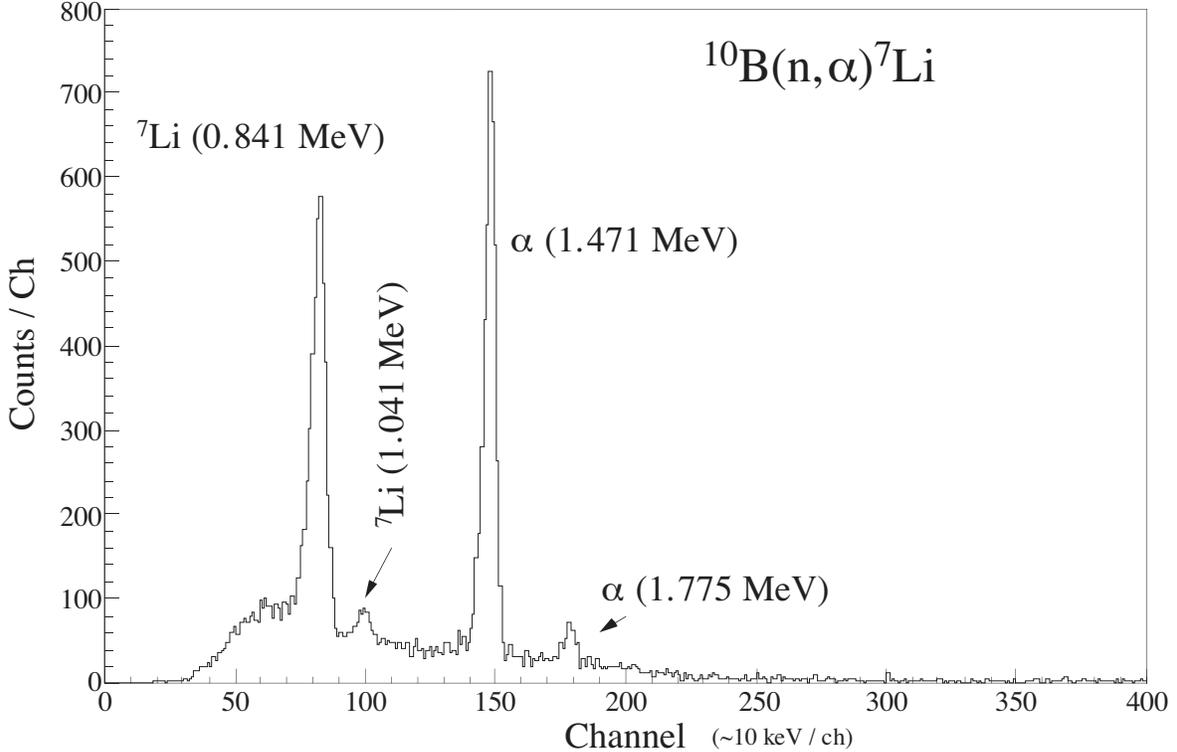}
 \caption{\label{10B}Typical $^{10}B(n,\alpha)^7Li$ calibration spectrum used in this study.}
\end{figure}

The energy calibration of the $\alpha$-detectors was performed using
$\alpha$-particles from five different standard calibration sources: $^{148}Gd,\  ^{241}Am,\ 
^{208}Po,\  ^{209}Po, \ ^{227}Ac$ and $\alpha$-particles as well as $^7Li$ emitted in the
$^{10}B(n,\alpha)^7Li$ reaction, yielding eleven $\alpha$-particle
energies from 1.471 MeV to 7.386 MeV. Energy loss in these calibration sources and
the $^{10}B(n,\alpha)^7Li$ source were negligible and did not affect the energy calibration. 
A typical $^{10}B(n,\alpha)^7Li$ calibration spectrum is shown in Fig.~\ref{10B}. 
In addition the detectors were implanted
with small amounts of Polonium and Actinium
leading to a continuous online energy calibration throughout the
experiment. These online calibration lines did not perturb the time of flight coincidence spectra.

\begin{figure}
 \includegraphics[height=4in]{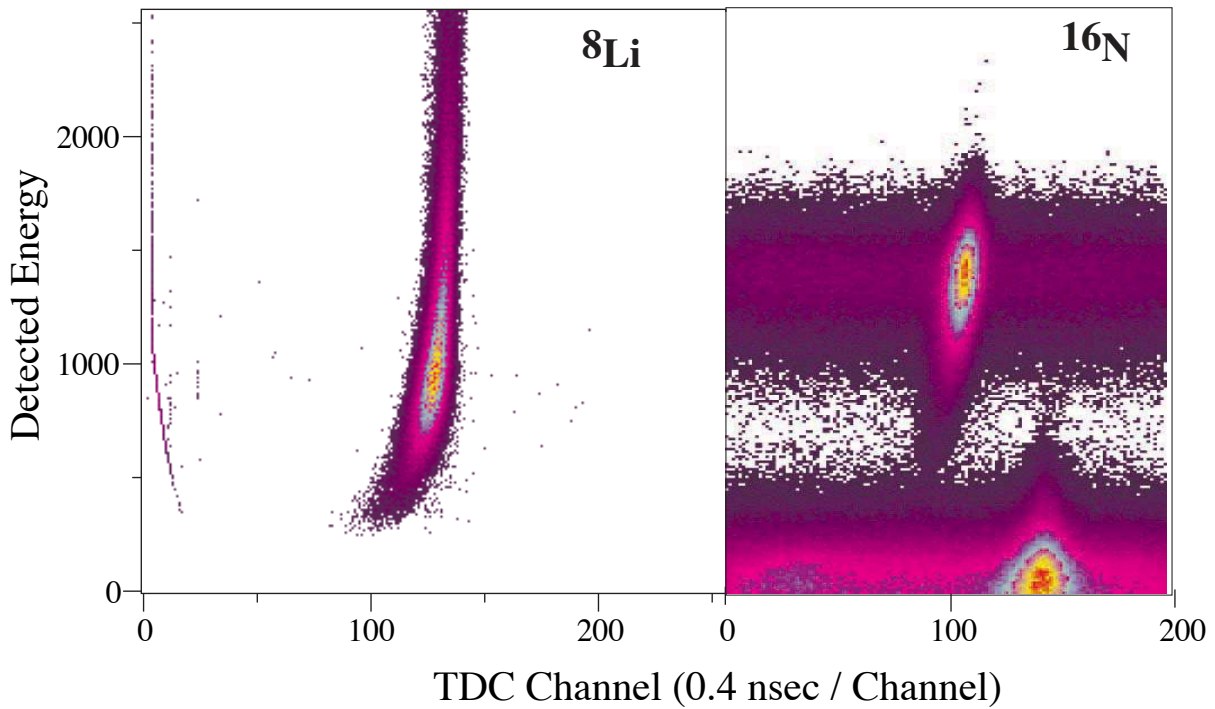}
 \caption{\label{2D}Typical two dimensional energy {\em vs.} time of flight 
 spectra for $^8Li$ and $^{16}N$ data.}
\end{figure}

\begin{figure}
 \includegraphics[height=6in]{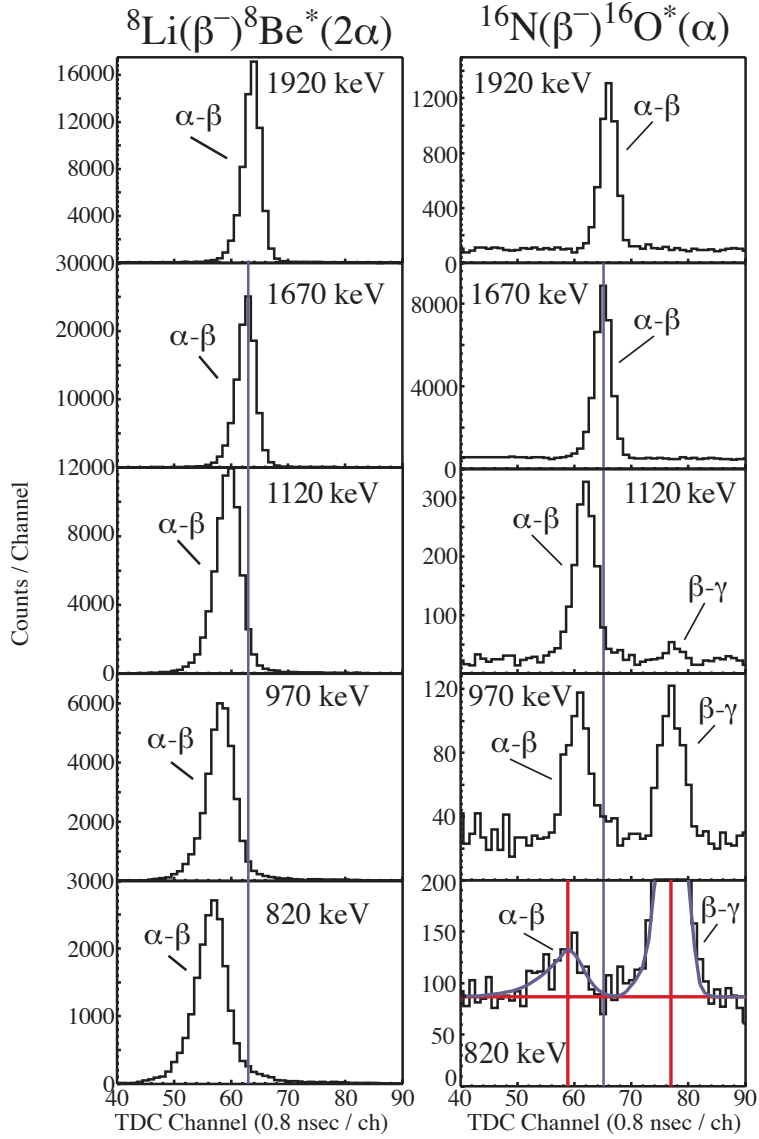}
 \caption{\label{slice}Typical $50\ keV$ wide slices from the 2 dimensional $^8Li$
  and $^{16}N$ energy vs time plots, see Fig.~\ref{2D}, projected upon the time axis. The slices are 
  labeled by the high energy end of the slice from which an effective center of mass 
  energy is determined using the known energy loss. The $\alpha$-$\beta$ 
coincidence peak is well separated from the $\beta$-$\gamma$ peak at low energies, 
and due to the kinematics, is also distinguishable from signals caused by partial
 charge collection in the $\alpha$-detector.  The $1670\ keV$ slice is at the central
 peak for $^{16}N$ and the line drawn through it (which represents the expected time of flight 
 location of a low energy tail) is clearly separated from the 
$\alpha$-$\beta$ coincidence peak at lower energies.  The $^8Li$ slices are located at 
slightly lower channels due to the different leading edge threshold of the higher 
energy $\beta$-particles in the Lithium decay.  }
\end{figure}

\begin{figure}
 \includegraphics[height=4in]{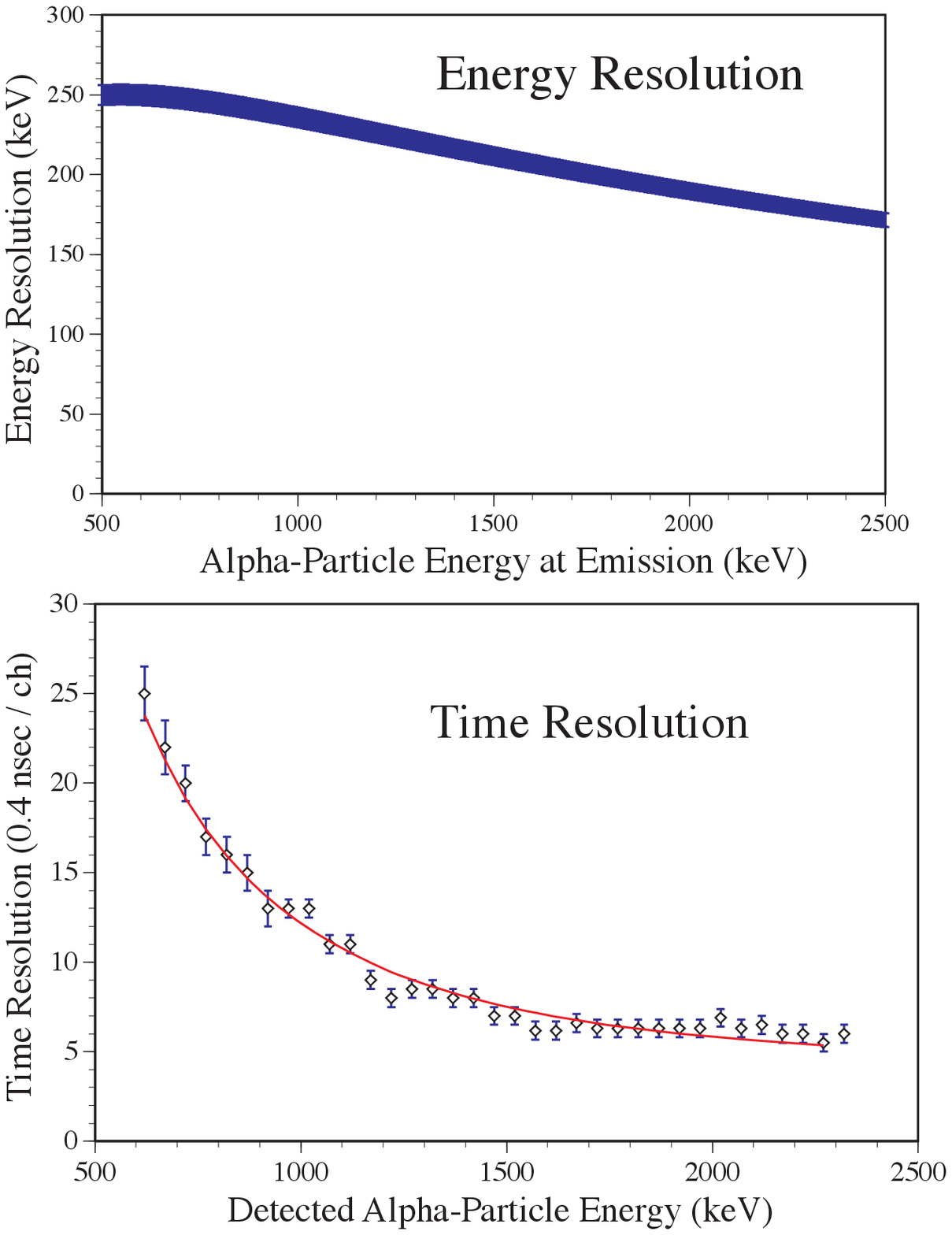}
 \caption{\label{res}The energy and time resolution (FWHM)
 as a function of energy. The energy resolution is determined by 
 the thickness of the aluminum catcher foil in which the 
$^{16}N$ is embedded.  The time resolution 
is based on measurements from the decay of $^8Li$.  }
\end{figure}

\section{Data Analysis}

The detector timing was calibrated using $22\ MeV$ $^7Li$ beams
and the $^2H(^7Li,^8Li)^1H$ reaction.  The $^8Li$ undergoes $\beta$-delayed $\alpha$-particle emission with a lifetime just under $1\ s$.  The measured $^8Li$ spectra were
plotted in 108 (one for each $\alpha$-$\beta$ detector pair) two-dimensional
histograms with time along the $x$-axis and detected $\alpha$-particle
energy along the $y$-axis, as shown in Fig.~\ref{2D}.  
Each $^8Li$ histogram was initially cut into sixteen
$100\ keV$ wide slices which were projected onto the time axis, similar to 
the (50 keV) slices shown in Fig.~\ref{slice}; the
resulting 1728 spectra were each fit using the Oak Ridge data analysis
program SAMGR with skewed gaussians to determine the time of flight (TOF) peak positions and
shapes, see Fig.~\ref{slice}.  For each detector pair, these TOF centroids were fit
as a function of energy; $Centroid = \sqrt{m_\alpha / 2\times Energy} *d + B$ where d is the
distance from the $\alpha$-particle source to the detector, as shown in Fig.~\ref{res}. 
The fitted values of d were consistant with the
distances measured directly (about $8.3\ cm$). The obtained time of 
flight parameters were used for matching the $^{16}N$ spectra. The time of flight 
resolution was determined from these $^8Li$ spectra and the energy resolution was 
determined by using ZieglerÕs formulae \cite{Zi92} with the known 
effective thickness measured {\em in situ} using a $^{148}Gd$ source. The time and 
energy resolutions are shown in Fig.~\ref{res}.

A $\beta$-delayed $\alpha$-particle emission spectrum of $^{16}N$ was obtained with approximately 1.3 million $\alpha$-particles in singles using the following procedure. 
After matching the spectra using the results of the 
$^8Li$ calibration as discussed above, the {$^{16}N$} spectra 
were combined into one two dimensional histogram, as shown in 
Fig~\ref{2D}.  This summed two dimensional spectrum was
then cut into $50\ keV$ wide slices which were individually analysed using
the Oak Ridge software, see Fig.~\ref{slice}, and corrected for
the measured $\beta$-particle efficiency of the $\beta$-array.   One of the $\beta$-detectors failed
during the experiment and was not used in the analysis.  For purely geometric reasons approximately
one third of the remaining detector pairings were not used in our analysis, as
their timing resolutions were insufficient to separate out the background (discussed below).

The effective energy of the emerging $\alpha$-particles for each data point was calculated 
using the expected variation of the yield over the energy 
width of the catcher foil for each slice. The effective center of mass energy was calculated
and is listed as $E_{cm}$ in Table 1. Note that due 
to fast variation in the yield, the effective $\alpha$-particle energy is not the one due to $\alpha$-particles 
emitted from the center of the catcher foil.

The measured spectral line shape was corrected for distortions caused by the
variability of our time and energy resolutions, see Fig.~\ref{res}. 
For a spectrum constant in energy, the yield measured at each point in that spectrum
is directly proportional to the energy integration interval.  This conclusion holds
for a spectrum in any physical variable.  In the case of this experiment, the data are
integrated over both time and energy with the integration intervals being the time and energy 
resolutions.  The fact that these vary considerably over the energy range of the detected $\alpha$-particles
causes a significant distortion in the line shape. The data were normalized by dividing by the effective
integration intervals, {\em i.e.} the time and energy resolutions as shown in Fig.~\ref{res}, 
to correct this distortion. The energy resolution of the experiment
(the thickness of the aluminum catcher foils) is considerably larger than the
intrinsic resolution of the SSB detectors. These resolutions are based
on the measured time resolutions from the $^8$Li data and the
measured thickness of the catcher foils \cite{France} determined by using ZieglerÕs formulae \cite{Zi92} upon the known 
effective thickness measured {\em in situ} using a $^{148}Gd$ source. The final spectrum, with these resolutions
divided out, is shown in Fig.~\ref{spectrum} and listed in tabular form in Table 1.

There are two principle sources of background in this experiment.  The
first is $\beta$-$\gamma$ coincidences; these occur when a $\beta$-particle is detected
in an $\alpha$-detector in coincidence with a $\gamma$-ray detected in a
$\beta$-detector.  Most of these $\beta$-$\gamma$ coincidences arise from activated
$^{28}Al$ created by neutron capture on the aluminum capture foils.
The second source of background is due to partial charge collection in the 
SSB.  In both cases the background coincidence is well separated from the data due
to the fast timing requirements, as shown in Fig~\ref{slice}.  

\underline{Table 1:} The currently measured Yale(96) data.

\begin{tabularx}{\linewidth}{|>{\setlength{\hsize}{.5\hsize}}X|%
>{\setlength{\hsize}{1.5\hsize}}X|%
>{\setlength{\hsize}{.5\hsize}}X|%
>{\setlength{\hsize}{1.5\hsize}}X|%
>{\setlength{\hsize}{.5\hsize}}X|%
>{\setlength{\hsize}{1.5\hsize}}X|%
>{\setlength{\hsize}{.5\hsize}}X|%
>{\setlength{\hsize}{1.5\hsize}}X|}
\hline
$E_{cm}$ & cts./ch &
$E_{cm}$ & cts./ch &
$E_{cm}$ & cts./ch &
$E_{cm}$ & cts./ch \\ \hline

975&46.1(175)&
1563&517.0(313)&
2149&19856.7(14335)&
2739&5241.7(4095)
\\ \hline
1040&67.5(104)&
1628&799.5(461)&
2215&26832.3(19323)&
2804&3082.4(2593)
\\ \hline
1105&72.5(114)&
1693&1227.0(683)&
2280&33244.0(23917)&
2869&1700.9(1584)
\\ \hline
1171&63.6(95)&
1758&1961.1(1500)&
2345&36684.7(26388)&
2936&958.3(1238)
\\ \hline
1236&85.1(95)&
1823&2907.9(2183)&
2412&35711.0(25700)&
3001&479.1(1684)
\\ \hline
1301&75.3(84)&
1889&4382.8(3244)&
2477&29550.8(21319)&
3067&238.0(939)
\\ \hline
1367&135.2(109)&
1853&6725.9(4928)&
2542&21181.2(15364)&
3132&123.5(906)
\\ \hline
1432&207.1(147)&
2019&9778.2(7116)&
2608&13623.0(9989)&
&
\\ \hline
1497&315.3(206)&
2084&14568.2(10545)&
2673&8626.0(6462)&
&
\\ \hline

\end{tabularx}

\section{Discussion}

In addition to our experiment there were three
high-statistics data sets for the $\beta$-delayed $\alpha$-particle emission of $^{16}$N.  
The first (containing approximately 32 million events) was taken in
Mainz, Germany by K. Neubeck {\em et al.} \cite{Ne74}, in a
successful measurement of
the $1.2825\ MeV$ $\alpha$-particles from the parity violating $\alpha$-particle decay
of the $8.8719\ MeV\ 2^-$ state in $^{16}O$.  The second (containing
approximately 1.25 million events) was taken in the TRIUMF lab in Canada
\cite{Az94}.  The third (containing approximately 0.1 million events)
was taken at the University of Washington in Seattle, Washington by Z. Zhao {\em et al.} \cite{Zh95}.

\begin{figure}
 \includegraphics[height=6in]{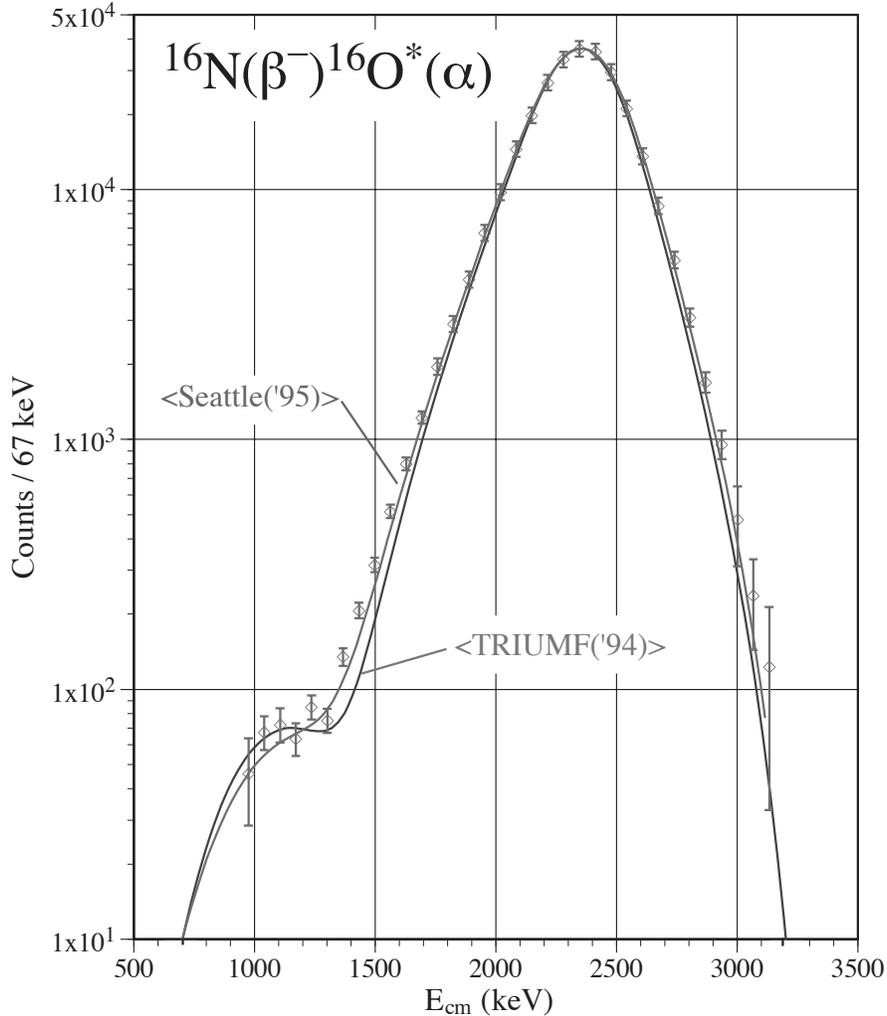}
 \caption{\label{spectrum} The $\beta$-delayed $\alpha$-particle emission spectrum of $^{16}N$
 obtained in this work. The TRIUMF and Seattle R-Matrix fitted curves \cite{Az94,Zh95} averaged
 over the energy resolution of our experiment (shown in Fig.~\ref{res}) are compared to our data.  }
\end{figure}

Due to the thickness of our catcher foils, the inherent energy resolution of
our experiment is significantly poorer than that of the previous experiments.
Thus, in order to do a proper comparison, the previous data sets must
be averaged over our variable energy resolution.  To do this,  the published R-Matrix fitted curves
from the Seattle and TRIUMF data sets were used to minimize end-effects in the 
averaging process.  The use of the R-matrix fit curve is
appropriate as the fitted curves reproduce the data quite well
\cite{Zh95,Az94} and it allows us to extend beyond the region of measured data for a 
meaningful averaging.

\begin{figure}
 \includegraphics[height=6in]{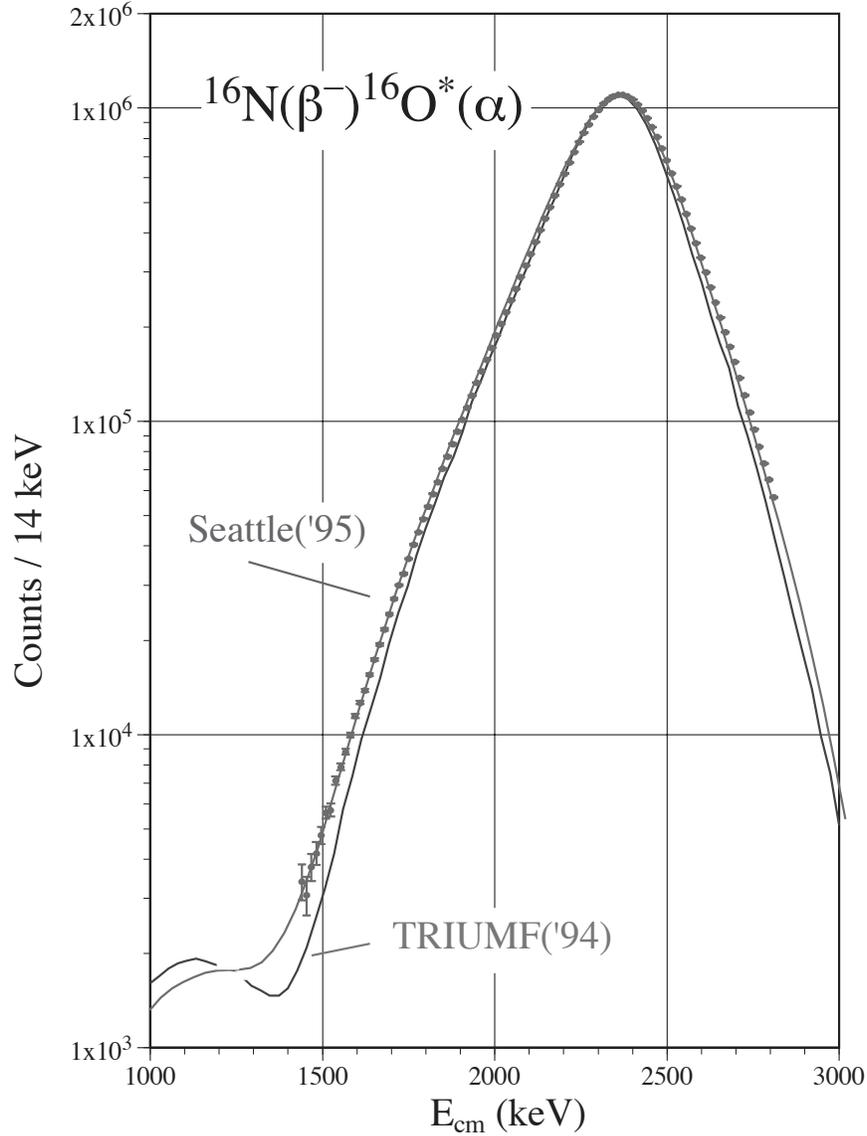}
 \caption{\label{Mainz}TRIUMF and Seattle R-Matrix fitted  curves  
  \cite{Az94, Zh95} compared with the Mainz data
  \cite{Wa69, Wa70,Ne74, Barker}. }
\end{figure}

In Fig~\ref{spectrum} we show a comparison of our data with the averaged fitted curves of
Seattle and TRIUMF.  Our data agree fairly well with the Seattle averaged fit curve
with a $\chi ^2$ per data point of 1.4, while disagreeing with the TRIUMF
averaged fit curve with a $\chi ^2$ per data point of 7.2.  In Fig.~\ref{Mainz} we show a
comparison of the Mainz data with the Seattle and TRIUMF fit  curves. These three data 
sets were measured with comparable energy resolution and in this comparison there is no 
need to employ energy averaging.  The
Mainz data also agrees with the Seattle fit curve with a $\chi ^2$ per data
point of 2.5, while badly disagreeing with the TRIUMF fit curve with a $\chi ^2$
per data point of 123.

The disagreement with the TRIUMF data is manifestly due to a difference in the
width of the primary peak, and not in the height of the secondary low energy peak. The agreement 
on the height of the low energy peak in and of itself demonstrates the viability of the time of 
flight method used in this experiment and negates a claim of a low energy tail in our data. 
The disagreement could quite possibly be due to over subtraction of $^{18}N$ contamination 
in the TRIUMF data.

In the case of $\beta$-delayed $\alpha$-particle emission 
of  $^{16}N$ the R-matrix fit includes  an additional partial wave, the f-wave, that affects the 
$^{16}N$ results, while not playing any role in the astrophysical  
$^{12}C(\alpha ,\gamma )^{16}O$
reaction.  The f-wave component is determined primarily through the depth of the
interference minimum around 1.4 MeV, which is also dependent upon the 
width of the primary peak. The disagreement in the region of the interference minimum 
around 1.4 MeV is sufficient to change the f-wave component and leads to imprecise 
determination of the p-wave contribution to the  $^{16}N$ spectrum.

\section{Conclusion}

We report on our improved measurement of the $\beta$-delayed $\alpha$-particle emission of $^{16}N$ with results that are in 
agreement with the old Mainz data and the new unpublished Seattle data but in 
disagreement with the TRIUMF data.  The disagreements are entirely caused by
differences in the width of primary peak and the depth of the interference minimum.  
Without precise knowledge of this minimum, the f-wave component cannot be 
determined with high accuracy and the corresponding p-wave spectrum cannot be 
extracted with high precision.

\newpage

\section{Appendix}

We include numerical values of other measured $\beta$-delayed $\alpha$-particle emission data sets:

\underline{Table A.1:} Mainz(Ô71)  data set  \cite{Wa69, Wa70,Ne74, Barker} with $\beta$ background
subtracted.

\begin{tabularx}{\linewidth}{|>{\setlength{\hsize}{.75\hsize}}X|%
>{\setlength{\hsize}{1.25\hsize}}X|%
>{\setlength{\hsize}{.75\hsize}}X|%
>{\setlength{\hsize}{1.25\hsize}}X|%
>{\setlength{\hsize}{.75\hsize}}X|%
>{\setlength{\hsize}{1.25\hsize}}X|%
>{\setlength{\hsize}{.75\hsize}}X|%
>{\setlength{\hsize}{1.25\hsize}}X|}

 \hline
$E_{cm}$ & cts./14 keV &
$E_{cm}$ & cts./14 keV &
$E_{cm}$ & cts./14 keV &
$E_{cm}$ & cts./14 keV \\ \hline

1440 &3405(450) &
1793 & 48833(221) &
2146&444856(667) &
2499&682050(826) 
\\ \hline
1454 & 3077(433) &
1807 & 53503(231)  &
2160&483692(695) &
2514&620224(788)
 \\ \hline
1468 & 4188(346) &
1821 & 58701(242) &
2174&526321(725) &
2528&563602(751)
 \\ \hline
1482&4188(346)&
1835&64079(253) &
2189&571892(756) &
2542&510728(715)
 \\ \hline
1496 & 4772(303) &
1849&70676(266) &
2203&618339(786) &
2556&459941(678)
 \\ \hline
1510 & 5642(256) &
1863&77366(278) &
2217&670463(819) &
2570&413566(643)
 \\ \hline
1524 & 5747(266) &
1878&84753(291) &
2231&724409(851) &
2584&371361(609)
 \\ \hline
1538 & 7140(213) &
1892&92990(305) &
2245&780412(883) &
2598&333038(577)
 \\ \hline
1553 & 7894(207) &
1906&101301(318) &
2259&835509(914) &
2613&299558(547)
 \\ \hline
1567 & 8809(187) &
1920&110729(333) &
2273&887989(942) &
2627&268149(518)
 \\ \hline
1581 & 9956(177) &
1934&120862(348) &
2287&940716(970) &
2641&240195(490)
 \\ \hline
1595 & 11465(166) &
1948&133076(365) &
2302&990074(995) &
2655&214700(463)
 \\ \hline
1609 & 12659(163) &
1962&144569(380) &
2316&1035270(1017) &
2669&192662(439)
 \\ \hline
1623 & 13843(165) &
1977&157468(397) &
2330&1067060(1033) &
2683&173030(416)
 \\ \hline
1637 & 15536(164) &
1991&171864(415) &
2344&1093030(1045) &
2697&154921(394)
 \\ \hline
1651 & 17367(166) &
2005&188071(434) &
2358&1102330(1050) &
2711&137601(371)
 \\ \hline
1666 & 19393(167) &
2019&204918(452) &
2372&1103420(1050) &
2726&121153(348)
 \\ \hline
1680 & 21678(172) &
2033&223316(473) &
2386&1090340(1044) &
2740&106848(327)
 \\ \hline
1694 & 24254(177) &
2047&243641(494) &
2401&1065430(1032) &
2754&94373(307)
 \\ \hline
1708 & 27143(184) &
2061&264433(514) &
2415&1027410(1014) &
2768&83053(288)
 \\ \hline
1722 & 30005(185) &
2075&289202(538) &
2429&983372(992) &
2782&73408(271)
 \\ \hline
1736 & 32645(195) &
2090&314792(561) &
2443&928649(964) &
2796&65206(255)
 \\ \hline
1750 & 36425(191) &
2104&342694(585) &
2457&869743(933) &
2810&57211(239)
 \\ \hline
1765 & 40454(201) &
2118&374401(612) &
2471&808732(899) &
&
 \\ \hline
1779 & 44290(210) &
2132&407842(639) &
2485&743730(862) &
&
 \\ \hline
\end{tabularx}

\newpage

\underline{Table A.2:} Seattle(Ô95) data set \cite{Zh95} as listed (by permission) in Ref. \cite{France}.

\begin{tabularx}{\linewidth}{|>{\setlength{\hsize}{.75\hsize}}X|%
>{\setlength{\hsize}{1.25\hsize}}X|%
>{\setlength{\hsize}{.75\hsize}}X|%
>{\setlength{\hsize}{1.25\hsize}}X|%
>{\setlength{\hsize}{.75\hsize}}X|%
>{\setlength{\hsize}{1.25\hsize}}X|%
>{\setlength{\hsize}{.75\hsize}}X|%
>{\setlength{\hsize}{1.25\hsize}}X|}

 \hline
$E_{cm}$ & cts./27 keV &
$E_{cm}$ & cts./27 keV &
$E_{cm}$ & cts./27 keV &
$E_{cm}$ & cts./27 keV \\ \hline

835&8(3)&
1454&25(5)&
2073&2141(46)&
2691&1142(34)
\\ \hline
868&13(4)&
1486&28(5)&
2105&2706(52)&
2724&821(29)
\\ \hline
900&16(4)&
1519&55(7)&
2138&3214(57)&
2757&643(25)
\\ \hline
933&10(3)&
1551&68(8)&
2170&3964(63)&
2789&466(22)
\\ \hline
965&12(3)&
1584&64(8)&
2203&4801(69)&
2822&395(20)
\\ \hline
998&11(3)&
1617&111(11)&
2235&5681(75)&
2854&256(16)
\\ \hline
1030&13(4)&
1649&120(11)&
2268&6533(81)&
2887&188(14)
\\ \hline
1063&12(3)&
1682&161(13)&
2301&7348(86)&
2919&118(11)
\\ \hline
1096&20(4)&
1714&232(15)&
2333&7870(89)&
2952&111(11)
\\ \hline
1128&10(3)&
1747&236(15)&
2366&8056(90)&
2985&72(8)
\\ \hline
1161&21(5)&
1779&352(19)&
2398&7777(88)&
3017&43(7)
\\ \hline
1193&7(3)&
1812&392(20)&
2431&7083(84)&
3050&34(6)
\\ \hline
1226&16(4)&
1845&546(23)&
2463&6000(78)&
3082&17(4)
\\ \hline
1258&13(4)&
1877&675(26)&
2496&4851(70)&
3115&10(3)
\\ \hline
1291&9(3)&
1910&862(29)&
2529&4038(64)&
3147&6(2)
\\ \hline
1323&15(4)&
1942&979(31)&
2561&3180(56)&
3180&1(1)
\\ \hline
1356&14(4)&
1975&1241(35)&
2594&2430(49)&
3212&0(1)
\\ \hline
1389&17(4)&
2007&1468(38)&
2627&1894(44)&
3245&1(1)
\\ \hline
1421&16(4)&
2040&1753(42)&
2659&1503(39)&
&
\\ \hline

\end{tabularx}

\end{document}